\documentclass[aps,prl, amsmath, showpacs, preprintnumbers,superscriptaddress, twocolumn, amssymb]{revtex4}

\usepackage{graphicx}
\usepackage{mathptmx, textcomp}
\usepackage[latin1]{inputenc}
\usepackage{tabularx}
\usepackage[usenames]{color}
\usepackage[colorlinks,urlcolor=blue,citecolor=blue,linkcolor=blue]{hyperref}

\begin{document}
\title{Universality of the Three-Body Parameter for Efimov States in Ultracold Cesium}
\author{M. Berninger}
\author{A. Zenesini}
\author{B. Huang}
\author{W. Harm}
\author{H.-C. N\"{a}gerl}
\author{F. Ferlaino}
\affiliation{Institut f\"ur Experimentalphysik and Zentrum f\"ur Quantenphysik, Universit\"at
 Innsbruck, 
 6020 Innsbruck, Austria}
\author{R. Grimm}
\affiliation{Institut f\"ur Experimentalphysik and Zentrum f\"ur Quantenphysik, Universit\"at
 Innsbruck, 
 6020 Innsbruck, Austria}
\affiliation{Institut f\"ur Quantenoptik und Quanteninformation,
 \"Osterreichische Akademie der Wissenschaften, 6020 Innsbruck,
 Austria}
\author{P. S. Julienne}
\affiliation{Joint Quantum Institute, NIST and the University of Maryland, Gaithersburg, Maryland 20899-8423, USA}
\author{J. M. Hutson}
\affiliation{Department of Chemistry, Durham University, South Road, Durham, DH1 3LE, United Kingdom}

\date{\today}

\begin{abstract}
We report on the observation of triatomic Efimov resonances in an ultracold gas of cesium atoms. Exploiting the wide tunability of interactions resulting from three broad Feshbach resonances in the same spin channel, we measure magnetic-field dependent three-body recombination loss. The positions of the loss resonances yield corresponding values for the three-body parameter, which in universal few-body physics is required to describe three-body phenomena and in particular to fix the spectrum of Efimov states. Our observations show a robust universal behavior with a three-body parameter that stays essentially constant.
\end{abstract}

\maketitle

The concept of universality manifests itself in the fact that different physical systems can exhibit basically the same behavior, even if the relevant energy and length scales differ by many orders of magnitude \cite{Braaten2006uif}.
Universality thus allows us to understand in the same theoretical framework physical situations that at first glance seem completely different.
In ultracold atomic collisions, the universal regime is realized when the $s$-wave scattering length $a$, characterizing the two-body interaction in the zero-energy limit, is much larger than the characteristic range of the interaction potential. Then the essential properties of the two-body system such as the binding energy of the most weakly bound dimer state and the dominating part of the two-body wave function can simply be described in terms of $a$, independent of any other system-dependent parameters. In the three-body sector, the description of a universal system requires an additional parameter, which incorporates all relevant short-range interactions not already included in $a$. In few-body physics, this important quantity is commonly referred to as the {three-body parameter} (3BP).

In Efimov's famous scenario \cite{Efimov1970ela,Braaten2006uif}, the infinite ladder of three-body bound states follows a discrete scaling invariance, which determines the relative energy spectrum of the states. The 3BP fixes the starting point of the ladder and thus the absolute energies of all states. The parameter enters the theoretical description as a short-range boundary condition for the three-body wave function in real space or as a high-frequency cut-off in momentum space. To determine the 3BP from theory would require precise knowledge of both the two-body interactions and the genuine three-body interactions at short range. In real systems, this is extremely difficult and the 3BP needs to be determined experimentally through the observation of few-body features such as Efimov resonances.

In the last few years, ultracold atomic systems have opened up the possibility to explore Efimov's scenario experimentally and to test further predictions of universal theory \cite{Kraemer2006efe, Knoop2009ooa, Zaccanti2009ooa, Pollack2009uit, Gross2009oou, Huckans2009tbr, Ottenstein2008cso, Williams2009efa, Barontini2009ooh, Gross2010nsi, Lompe2010ads, Nakajima2010nea}. The key ingredient of such experiments is the possibility to control $a$ 
by an external magnetic field $B$ via the Feshbach resonance phenomenon \cite{Chin2010fri}. This naturally leads to the important question whether the 3BP remains constant or whether it is affected by the magnetic tuning, in particular when different Feshbach resonances are involved.

The current status of theoretical and experimental research does not provide a conclusive picture on possible variations of the 3BP. A theoretical study \cite{Dincao2009tsr} points to strong possible variations when different two-body resonances are exploited in the same system, and even suggests a change of the 3BP on the two sides of a zero crossing of the scattering length. Other theoretical papers point to the importance of the particular character of the Feshbach resonance \cite{Chin2010fri}. While closed-channel dominated (``narrow'') resonances involve an additional length scale that may fix the 3BP \cite{Petrov2004tbp,Massignan2008esn,Wang2010utb}, the case of entrance-channel dominated (``broad'') resonances leaves the 3BP in principle open. However, predictions based on two-body scattering properties exist that apparently fix the 3BP for broad resonances as well \cite{Lee2007ete, Jonalasinio2010tru, Naidon2011tee}. The available experimental observations provide only fragmentary information. The first observation of Efimov physics in an ultracold Cs gas \cite{Kraemer2006efe} is consistent with the assumption of a constant 3BP on both sides of a zero crossing. A later observation on $^{39}$K \cite{Zaccanti2009ooa} indicated different values of the 3BP on both sides of a Feshbach resonance. A similar conclusion was drawn from experiments on $^7$Li \cite{Pollack2009uit}, but other experiments on $^7$Li showed universal behavior with a constant 3BP for the whole tuning range of a single resonance \cite{Gross2009oou} and for another spin channel \cite{Gross2010nsi}. Besides these observations on bosonic systems, experiments on fermionic gases of $^6$Li \cite{Ottenstein2008cso, Huckans2009tbr, Williams2009efa} can be interpreted based on a constant 3BP \cite{Wenz2009uti}. A recent experiment on $^6$Li, however, indicates small variations of the 3BP \cite{Nakajima2011moa}.

\begin{figure}
 \includegraphics[width=8.5cm] {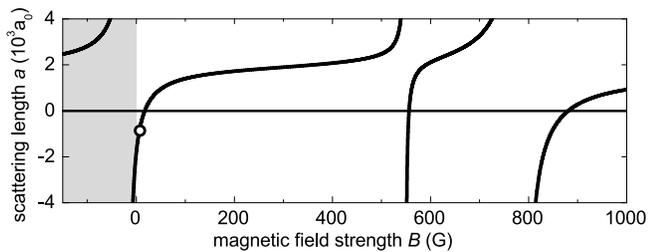}
 \caption{Illustration of the three broad $s$-wave Feshbach resonances for Cs in the absolute atomic ground state $|F\!=\!3, m_F\!=\!3 \rangle$. The open circle corresponds to the previous observation of a triatomic Efimov resonance at $7.6$\,G \cite{Kraemer2006efe}. The many narrow Feshbach resonances resulting from $d$- and $g$-wave molecular states \cite{Chin2004pfs} are not shown for the sake of clarity. The region with $B<0$ corresponds to the state $|F\!=\!3, m_F\!=\!-3 \rangle$, which is not stable against two-body decay. Scattering lengths are given in units of Bohr's radius $a_0$.}
 \label{Bscan}
\end{figure}

In the present work, we investigate universality in an ultracold gas of Cs atoms, which offers several broad Feshbach resonances in the same spin channel and thus offers unique possibilities to test for variations of the 3BP.
In the lowest hyperfine and Zeeman sublevel $|F\!=\!3, m_F\!=\!3 \rangle$, Cs features a variety of broad and narrow Feshbach resonances in combination with a large background scattering length \cite{Chin2004pfs}. Of particular interest are three broad $s$-wave Feshbach resonances in the range up to 1000\,G \cite{gauss}, with poles near $-10$~G, $550$~G, and $800$~G \cite{Chin2004pfs, Lee2007ete, Chin2010fri}. The character of these three resonances is strongly entrance-channel dominated, as highlighted by the large values of their resonance strength parameter $s_{\rm res}$ \cite{Chin2010fri} of $560$, $170$, and $1470$, respectively. The resulting magnetic-field dependence $a(B)$ is illustrated in Fig.\,\ref{Bscan}.
In our previous work \cite{Kraemer2006efe,Knoop2009ooa} we have focussed on the low-field region up to $150$~G. After a major technical upgrade of our coil set-up, we are now in the position to apply magnetic fields $B$ of up to 1.4~kG with precise control down to the 20~mG uncertainty level and thus to explore the resonance regions at 550~G and 800~G \cite{BerningerPhD}.

Our ultracold sample consists of about $2 \times 10^4$ optically trapped $^{133}$Cs atoms, close to quantum degeneracy. The preparation is based on an all-optical cooling approach as presented in Refs.\,\cite{Kraemer2004opo, Mark2007sou}. The final stage of evaporative cooling is performed in a crossed-beam dipole trap (laser wavelength 1064\,nm)
and stopped shortly before Bose-Einstein condensation is reached. Finally, the trap is adiabatically recompressed to twice the initial potential depth to suppress further evaporation loss. At this point, the mean trap frequency is about 10\,Hz and the temperature is typically 15~nK.

Our experimental observable is the three-body loss coefficient $L_3$,
which in the framework of universal theory is conveniently expressed as $L_3\, = \,3 C(a) \frac{\hbar a^4}{m}$ \cite{Weber2003tbr},
$m$ denotes the atomic mass. The expression separates a log-periodic function $C(a)$ from the general $a^4$-scaling of three-body loss. For $a<0$, effective field theory \cite{Braaten2006uif} provides the analytic expression
\begin{equation}\label{equ:lossCoeff}
 C(a) = 4590\,\frac{\sinh(2\eta_-)}{\sin^2[s_0\ln(a /a_-)]+\sinh^2\eta_-},
\end{equation}
with $s_0 \approx 1.00624$ for identical bosons. The decay parameter $\eta_-$ is a non-universal quantity that depends on the deeply-bound molecular spectrum \cite{Wenz2009uti}. The scattering length $a_-$ marks the situation where an Efimov state intersects the three-atom threshold and the resulting triatomic Efimov resonance leads to a giant three-body loss feature. In the following, the quantity $a_-$ will serve us as the representation of the 3BP.


To measure $L_3$ we record the time evolution of the atom number after quickly (within 10\,ms) ramping $B$ from the evaporation to the target field strength.
We determine the atom number $N$ by absorption imaging. One-body decay, as caused by background collisions, is negligible under our experimental conditions. Furthermore, two-body decay is energetically suppressed in the atomic state used. We can therefore model the decay by $\dot{N}/N\,=\,-L_3 \langle n^2 \rangle$, where the brackets denote the spatial average weighted with the atomic density distribution $n$. Additional, weaker loss contributions caused by four-body recombination \cite{Ferlaino2009efu} can be described in terms of an effective $L_3$ \cite{Vonstecher2009sou}. For fitting the decay curves and extracting $L_3$ we use an analytic expression that takes into account the density decrease resulting from anti-evaporation heating \cite{Weber2003tbr}.

The experimental results are a function of $B$ whereas theory expresses $L_3$ as a function of $a$. It is thus crucial to have a reliable conversion function $a(B)$. We have obtained $a(B)$ from full coupled-channel calculations on a Cs-Cs potential obtained by least-squares fitting to extensive new measurements of binding energies, obtained by magnetic-field modulation spectroscopy \cite{Lange2009doa}, together with additional measurements of loss maxima and minima that occur at resonance poles and zero crossings. The new potential provides a much improved representation of the bound states and scattering across the whole range from low field to 1000 G. The experimental results and the procedures used to fit them will be described in a separate publication.

\begin{figure*}
\includegraphics[width=1.5 \columnwidth] {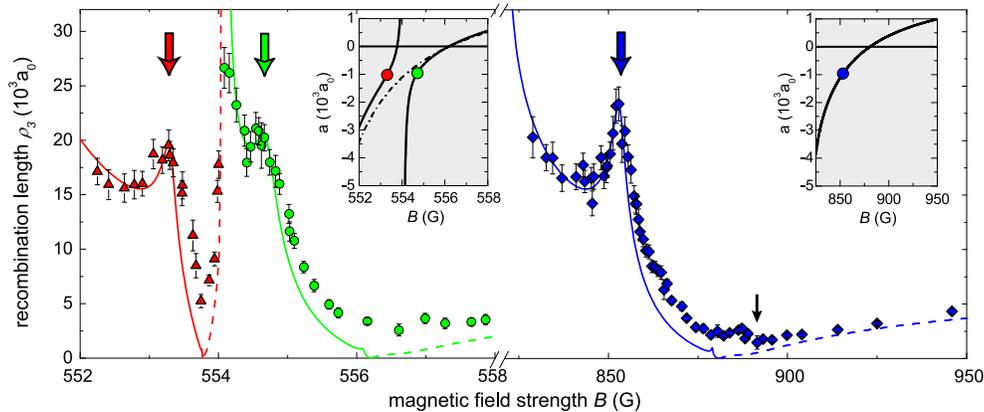}
\caption{(color online) Recombination loss in the vicinity of the high-field Feshbach resonances. The measured recombination length $\rho_3$ is shown for three different regions ($\blacktriangle$, $552\,\mathrm{G} < B < 554\,\mathrm{G}$;  $\bullet$, $554\,\mathrm{G} < B < 558\,\mathrm{G}$; $\blacklozenge$, $830\,\mathrm{G} < B < 950\,\mathrm{G}$), which are separated by the poles of different Feshbach resonances. The error bars indicate the statistical uncertainties. For all three regions, the solid lines represent independent fits to the data at negative $a$.
The dashed lines show the predictions of effective field theory for $a>0$ \cite{Braaten2006uif}, using the parameters obtained in the same region at $a<0$.
The insets show $a(B)$ (solid line, full calculation; dash-dot line, $s$-wave states only). The arrows in the main figure and the corresponding dots in the insets refer to the triatomic Efimov resonances. The small arrow indicates a recombination minimum.
}\label{EfimovMagneticField}
\end{figure*}

Figure~\ref{EfimovMagneticField} shows our experimental results on the magnetic-field dependent recombination loss near the two broad high-field Feshbach resonances (550\,G and 800\,G regions). For convenience we plot our data in terms of the recombination length $\rho_3\, = \,(2m L_3/(\sqrt{3}\hbar))^{1/4}$ \cite{Esry1999rot}.
The three filled arrows indicate three observed loss resonances that do not coincide with the poles of two-body resonances. We interpret these three features as triatomic Efimov resonances.

In the 800~G region, a single loss resonance shows up at 853\,G, which lies in the region of large negative values of $a$. We fit the $L_3$ data based on Eq.\ (\ref{equ:lossCoeff}) \cite{scalingfactor} and using the conversion function $a(B)$ described above. The fit generally reproduces the experimental data well, apart from a small background loss that apparently does not result from three-body recombination \cite{Lifetime}. For the 3BP the fit yields the resonance position of $a_- \, = \, -955(28) \, a_0$, where the given error includes all statistical errors. For the decay parameter the fit gives $\eta_- = 0.08(1)$.

For the 550\,G region, Fig.\ \ref{Bscan} suggests a qualitatively similar behavior as found in the 800\,G region. The experimental data, however, reveal a more complicated structure with three loss maxima and a pronounced minimum. This behavior is explained by a $g$-wave resonance (not shown in Fig.\ \ref{Bscan}) that overlaps with the broader $s$-wave resonance. We have thoroughly investigated this region by Feshbach spectroscopy.
These studies clearly identify the central maximum (554.06\,G) and the deep minimum (553.73\,G) as the pole and zero crossing of the $g$-wave resonance (see inset). With $s_{\rm res}=0.9$, this resonance is an intermediate case between closed-channel and entrance-channel dominated.

The $g$-wave resonance causes a splitting that produces two Efimov resonances instead of one in this region. This explains the upper and the lower loss maxima, which are found at 553.30(4)~G and 554.71(6)~G (arrows in Fig.\,\ref{EfimovMagneticField}). To determine the parameters of these Efimov resonances, we independently fit the two relevant regions of negative scattering length using Eq.\,(\ref{equ:lossCoeff}) \cite{scalingfactor}. This yields $a_-\, = \,-1029(58)\, a_0$ and $\,-957(80) \, a_0$ for the lower and the upper resonances, respectively.

\begin{figure}
\vspace{0mm}
 \includegraphics[width=7.5cm] {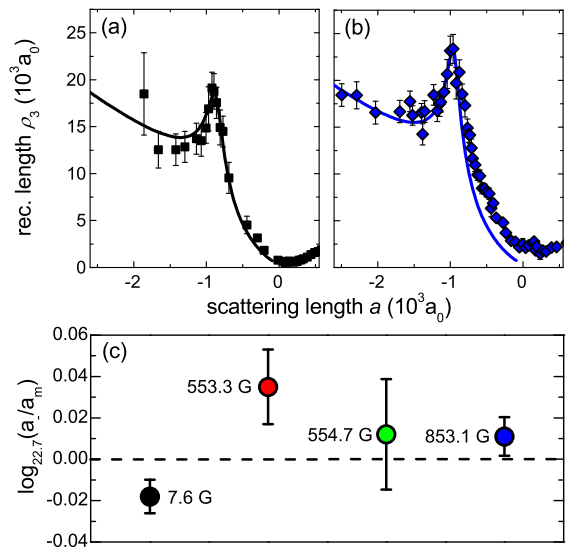}
 \vspace{0mm}
 \caption{(color online) Efimov resonances and the 3BP. In (a) and (b), we compare the resonance previously observed \cite{Kraemer2006efe} at 7.6\,G to the one found at 853\,G. In (c), we plot the 3BPs obtained for all four resonances measured in Cs. The dashed line corresponds to a mean value of $a_m = -921\,a_0$, calculated as a weighted average of the four different values. The logarithmic scale (to the basis of $22.7$) covers one tenth of the Efimov period.}

\label{EfimovResAll}
\end{figure}

\begin{table}
\begin{center}
\begin{tabularx}{7cm}{c c c c c c}
\hline	\hline		
    $B_{\rm res} $ (G) & $a_-$/$a_0$  & $\delta_1/a_0$& $\delta_2/a_0$ &$\delta_3/a_0$& $\eta_-$ \\
  \hline
  7.56(17)& $-872$(22) & 21 & 3 & 6& 0.10(3) \\

  553.30(4) & $-1029$(58) & 43 & 28 & 27 & 0.12(1) \\
  554.71(6) & $-957$(80) & 57 & 25 & 49& 0.19(2) \\

  853.07(56) & $-955$(28) & 27 & 1 & 4& 0.08(1) \\
\hline
\hline
\end{tabularx}
\end{center}
\caption{Parameters of the four triatomic Efimov resonances. The first and second column give the magnetic field values $B_{\rm res}$ at the resonance centers and the corresponding 3BPs together with their full statistical uncertainties. The individual error contributions $\delta_1$,  $\delta_2$, and $\delta_3$ refer to the statistical uncertainties from the fit 
to the $L_3$ data, from the determination of the magnetic field strength, and from the $a(B)$-conversion, respectively.} 
\label{table2}
\end{table}

We now compare all our observations on triatomic Efimov resonances in Cs. We also include the previous data of Ref.\ \cite{Kraemer2006efe} on the low-field resonance (7.6\,G), which we have refitted using our improved $a(B)$ conversion. The relevant parameters for the four observed Efimov resonances are given in Table \ref{table2}. Figure~\ref{EfimovResAll}(a) and (b) show the recombination data for the low-field resonance and the 853\,G resonance, using a convenient $\rho_3(a)$ representation. This comparison illustrates the striking similarity between both cases. For all four Efimov resonances, Fig.~\ref{EfimovResAll}(c) shows the 3BP on a logarithmic scale, which relates our results to the universal scaling factor 22.7. Note that the full scale is only one tenth of the Efimov period, i.e.\ a factor $22.7^\frac{1}{10} \approx 1.37$. The error bars indicate the corresponding uncertainties (one standard deviation), resulting from all statistical uncertainties \cite{systematics}. The data points somewhat scatter around an average value of about $-921$\,$a_0$ (dashed line) with small deviations that stay within a few percent of the Efimov period. Taking the uncertainties into account, our data are consistent with a constant 3BP for all four resonances. However, between the values determined for the two broad resonances at 7.6 and 853\,G we find a possible small aberration of about $2.5$ standard deviations. This may be accidental but it may also hint at a small change in the 3BP.

Let us briefly discuss our findings on further few-body observables. For $a>0$, three-body recombination minima are well known features related to Efimov physics \cite{Braaten2006uif, Kraemer2004opo, Zaccanti2009ooa, Pollack2009uit}. In the 800\,G region, we observe a minimum at $B=893(1)$\,G (small arrow in Fig.~\ref{EfimovMagneticField}), corresponding to $a = +270(30)\,a_0$, which is very similar to the minimum previously observed in the low-field region \cite{Kraemer2004opo} and consistent with a universal connection to the $a<0$ side. In general, however, these minima are difficult to access in Cs and dedicated experiments will be needed to provide stringent tests also for the $a>0$ side. Also atom-dimer resonances \cite{Knoop2009ooa, Zaccanti2009ooa, Pollack2009uit} have not yet been observed in the high-field region. Additional measurements in the 800\,G region (not shown) reveal a pair of four-body resonances at $865.4(5)$\,G and $855.0(2)$\,G, corresponding to scattering lengths of $-444(8)\,a_0$ and $-862(9)\,a_0$. This excellently fits to universal relations \cite{Vonstecher2009sou} and our previous observations at low magnetic fields \cite{Ferlaino2009efu}.

Our observations show that universality persists in a wide magnetic-field range across a series of Feshbach resonances in the same spin channel and that the 3BP shows only minor variations, if any. This rules out a scenario of large variations as suggested by the model calculations of Ref.\,\cite{Dincao2009tsr}.
The apparent fact that the relevant short-range physics is not substantially affected by the magnetic field may be connected to the strongly entrance-channel dominated character \cite{Chin2010fri} of the broad resonances in Cs. However, even the case of overlapping $s$- and $g$-wave Feshbach resonances, where the latter one has intermediate character, is found to exhibit universal behavior consistent with an essentially constant 3BP.
Our observation that universality is robust against passing through many poles and zero crossings of the scattering length also implies a strong argument in favor of a universal connection of both sides of a single Feshbach resonance. This supports conclusions from experiments on $^7$Li as reported in Refs.~\cite{Gross2009oou,Gross2010nsi}, in contrast to Ref.~\cite{Pollack2009uit} and related work on $^{39}$K \cite{Zaccanti2009ooa}.

With the present experimental data there is growing experimental evidence that theories based on low-energy two-body scattering and the near-threshold dimer states \cite{Lee2007ete, Jonalasinio2010tru, Naidon2011tee} can provide reasonable predictions for the 3BP without invoking genuine short-range three-body forces, which are known to be substantial for all the alkali metal trimers \cite{Soldan2003tbn}. We also stress a remarkable similarity \cite{Gross2009oou} between the Cs data and experimental results on both Li isotopes. When the 3BP is normalized to the mean scattering length $\bar{a}$ of the van der Waals potential \cite{Chin2010fri}, our actual Cs value $a_-/\bar{a} = -9.5(4)$ is remarkably close to corresponding values for $^7$Li \cite{Gross2010nsi,Pollack2009uit} and $^6$Li \cite{Ottenstein2008cso,Huckans2009tbr,Williams2009efa,Wenz2009uti}, which vary in the range between $-8$ and $-10$.

Universality in tunable atomic quantum gases near Feshbach resonances appears to be rather robust, but the understanding of the particular reasons and conditions remains a challenge to few-body theories.

We thank M.~Ueda, T.~Mukaiyama, P.~Naidon, and S.~Jochim for discussions. We acknowledge support by the Austrian Science Fund FWF within project P23106. A.Z.\ is supported within the Marie Curie Intra-European Program of the European Commission. P.S.J.\ and J.M.H.\ acknowledge support from an AFOSR MURI grant.


\begin{thebibliography}{38}
\bibitem[{gau()}]{gauss}
\bibinfo{note}{Units of gauss instead of the SI unit tesla ($1\,{\rm
  G}=10^{-4}\,{\rm T}$) are used to conform to conventional usage in this
  field.}
\bibitem[{sca()}]{scalingfactor}
\bibinfo{note}{The fits include an additional free scaling factor $\lambda$ to
  account for possible systematic errors in the number density. For the 800~G
  region, we obtain $\lambda=0.89$. For the lower and upper 550~G region, we
  obtain $\lambda=0.46$ and $\lambda=1.06$, respectively.}
\bibitem[{sys()}]{systematics}
\bibinfo{note}{Systematic uncertainties are not included in our error budget.
  They may result from model-dependent errors in the determination of the
  scattering length from binding energy and scattering data and from
  finite-temperature shifts. All these errors, however, stay well below the
  statistical uncertainties.}
\end{thebibliography}

\end{document}